\documentclass[aps,twocolumn]{revtex4}

\input epsf
\usepackage{amsmath}
\usepackage{amsfonts}
\usepackage{amssymb}%

\newcommand{\ME}[3]{\langle #1 | #2 | #3 \rangle}

\newcommand{\half}{\frac12}
\newcommand{\GeV}{{\rm\ GeV}}
\newcommand{\TeV}{{\rm\ TeV}}

\newcommand{\tr}{{\rm tr \ }}
\newcommand{\nonoindent}{}
\newcommand{\checker}[1]{#1_v}

\begin{document}

\title{
Echoes of a Hidden Valley at Hadron Colliders}

\author{Matthew J. Strassler
and Kathryn M. Zurek}
\affiliation{Department of Physics,
P.O Box 351560, University of Washington,
Seattle, WA 98195\\
}
\begin{abstract}{We consider examples of ``hidden-valley''
models, in which a new confining gauge group is added to the standard
model.  Such models often arise in string constructions, and elsewhere. 
The resulting (electrically-neutral) bound states can have low
masses and long lifetimes, and could be observed at the LHC and
Tevatron.  Production multiplicities are often large.  Final states
with heavy flavor are common; lepton pairs, displaced vertices and/or
missing energy are possible.  
Accounting for LEP
constraints, we find LHC production cross-sections typically in the
1-100 fb range, though they can be larger.  It is possible the Higgs
boson could be discovered at the Tevatron through rare decays to the
new particles.  }
\end{abstract}

\maketitle

As the era of the Large Hadron Collider (LHC) approaches, and the
Tevatron accumulates data, it is important to consider
well-motivated particle physics models that present novel and
challenging experimental signals.  There has been much recent work
along these lines \cite{split,NMSSM,Wells,WeirdHiggs,Bargeretal}.  

Here we discuss a class of ``hidden-valley'' models that
has received little study.  These are models in which the
standard model (SM) gauge group $G_{SM}$ is extended by a non-abelian group
$\checker G$.  All SM particles are neutral under $\checker G$, but there
are new light particles (``v-particles'') charged under $\checker G$ and
neutral under $G_{SM}$.  Higher dimension operators at the TeV scale
(induced perhaps by a $Z'$ or by a loop of heavy particles carrying
both $G_{SM}$ and $\checker G$ charges) allow interactions between SM
fields and the new light particles.  In this circumstance, the
v-particles are rarely produced at LEPI or LEPII, but may be
abundantly produced at the LHC, and perhaps even the Tevatron.
In a confining hidden-valley model,
%Confinement in the $\checker G$ sector then ensures that 
all
v-particles assemble themselves into $\checker G$-neutral ``v-hadrons.''
Some of the v-hadrons can then decay, again
via higher dimension operators, to gauge-invariant combinations of SM
particles, with observable lifetimes.  The diverse masses and
lifetimes of the v-hadrons, their multiplicities, and the
variety of possible final states make the v-phenomenology complex, and
sensitive to underlying parameters, such as v-quark masses.  

The hidden-valley scenario is consistent with data and is well
motivated; it arises in many top-down models, including string-theory
constructions \cite{stringexamples}.  It appears consistent with most
methods for solving the hierarchy problem (supersymmetry, little Higgs
models, TeV extra dimensions, Randall-Sundrum scenarios) so the
v-phenomenology outlined below may accompany more familiar physics
associated with each of these solutions.  However, the purpose of this
paper is not model-building, but rather to call attention to generic
signals which may not yet have been fully explored at the Tevatron or
in studies for the LHC.

As there is no clearly
``minimal'' hidden-valley model, we will present a simple theory 
exhibiting many phenomena typical of a v-sector, which include the following:
\newline $\bullet$  There are several long-lived v-hadrons, with
masses typically of order the v-confinement scale $\checker\Lambda$.
\newline $\bullet$  Some v-hadrons may be stable, providing dark matter 
candidates and missing energy signals, while others
decay to neutral combinations of SM particles.
\newline $\bullet$  Decay lifetimes can vary over many orders of magnitude;
v-hadrons may decay promptly, or produce a displaced
vertex anywhere in the detector, or typically decay outside
the detector.
\newline $\bullet$  Some v-hadrons decay preferentially to heavy flavor, while
others decay more democratically to $f\bar f$ final states
($f$ any SM fermion) or to $f\bar f$ plus 
another v-hadron; other final
states can include two or three gluons, $WW$ or $ZZ$.
\newline $\bullet$  V-hadron production multiplicities at the LHC may 
be large, especially if $\checker\Lambda \ll 1$ TeV.

%After discussing how these effects arise within our illustrative model,
%we will briefly mention another class of theories where
%these phenomena arise differently.

\begin{table}[b]
\begin{tabular}[c]{||c||c|c|c|c|c|c||c|c|c|c||c|c||}\hline
\ & $q_i$ &$\bar u_i$& $\bar d_i$ 
& $\ell_i$ & $e^+_i$ & $N_i$ & $U$ & $\bar U$ & $C$ & $\bar C$ & $H$ & $\phi$
\\ \hline %\hline
$SU(3)$ & ${\bf 3}$ & ${\bf \overline 3}$  & ${\bf \overline 3}$ 
&  ${\bf 1}$  & ${\bf  1}$ &  ${\bf 1}$  & ${\bf  1}$ 
&  ${\bf 1}$  & ${\bf  1}$ &  ${\bf 1}$  & ${\bf  1}$ &  ${\bf 1}$  
\\ \hline
$SU(2)$ & ${\bf 2}$ & ${\bf 1}$  & ${\bf  1}$ 
&  ${\bf 2}$  & ${\bf  1}$ &  ${\bf 1}$  & ${\bf  1}$ 
&  ${\bf 1}$  & ${\bf  1}$ &  ${\bf 1}$  & ${\bf  2}$ &  ${\bf 1}$  
\\ \hline
$U(1)_Y$ & $\frac16$ 
& $-\frac23$  & $\frac13$ 
& $-\frac12$ & $1$  & $0$ & $0$ & $0$ & $0$ & $0$ & $\frac12$ & $0$ 
 \\  \hline %\hline
$U(1)_\chi$ & $-\frac15$ & $-\frac15$  & $\frac35$ 
& $\frac35$ & $-\frac15$ & $-1$ 
%& $q_U$ & $2-q_U$ & $q_C$ & $2-q_C$ 
%& $q$ & $2-q$ & $2-q$ & $q$ 
& $q_+$ & $q_-$ & $-q_+$ & $-q_-$ 
& $\frac25$ & $2$
\\ \hline
$SU(\checker n)$ & ${\bf 1}$ & ${\bf 1}$  & ${\bf  1}$ 
&  ${\bf 1}$  & ${\bf  1}$ &  ${\bf 1}$  
&  ${\bf \checker n}$  & ${\bf  \overline{\checker n}}$ 
&  ${\bf \checker n}$  & ${\bf  \overline{\checker n}}$ 
& ${\bf  1}$ &  ${\bf 1}$  
\\ \hline
\end{tabular}
\caption{Charge assignments for the model before
removal of kinetic mixing;
$q_+ + q_- = -2$. 
}
\label{charges}
\end{table}

\nonoindent\underline{A Simple v-Model:} To the SM, we add a
$U(1)\times SU(\checker n)$ gauge group, with couplings $g'$ and $\checker
g$; we omit the special case $\checker n=2$.  
The $SU(\checker n)$ interaction confines
at the scale $1 \GeV< \checker\Lambda<$ 1 TeV.  
The $U(1)$ is broken by a scalar 
expectation value $\langle\phi\rangle$, giving a $Z'$ of mass $\sim 1-6 \TeV$.
We add two v-quark flavors $U,\bar U$ and $C, \bar C$ in the $\checker n$
and $\overline{\checker n}$ representation,
and three right-handed neutrinos $N_i$; all become
massive by coupling to $\phi$. 
%The $U$, $C$ and $N_i$
%All of these
%particles 
%become massive by coupling to $\phi$, which, combined
%with 
These masses and anomaly cancellation restrict
the $U(1)$ charges to those shown in table \ref{charges}, with
$q_+$ arbitrary and $q_+ + q_-=-2$. 
(We omit the special case $q_+=q_-=-1$.)  
For definiteness, we take the SM particles (and the $N_i$) to
be charged as under the usual $U(1)_\chi$ subgroup of $SO(10)$
that commutes with $G_{SM}$; see \cite{Langacker}.  
We impose a $Z_2$ symmetry
forbidding Dirac neutrino masses, making the $N_i$ 
stable (and dark matter candidates.)
Kinetic mixing between 
$U(1)_\chi$ and hypercharge, via the interaction $\half k F'_{\mu\nu}
F_Y^{\mu\nu}$, cannot be forbidden; we treat $k$
as a free parameter.  
We may
remove $k$ by a field redefinition, at the cost of
changing the charges.
{\it In our formulas below,
all factors of $g'$ and all $U(1)_\chi$ charges $Q_i$ are defined  after
removal of $k$ by field redefinition}; the new 
charges differ from those in Table \ref{charges} by
$Q_\chi=Q^{0}_\chi - (k g_1/g') Y$
and, as expected \cite{kmixing}, are no longer quantized.
Some other classic $U(1)$
groups, such as $B-L$, are obtained for
special values of $k$.

\begin{figure}[htbp]
  \begin{center}
    \leavevmode
     \epsfxsize=.45 \textwidth
     \hskip 0in \epsfbox{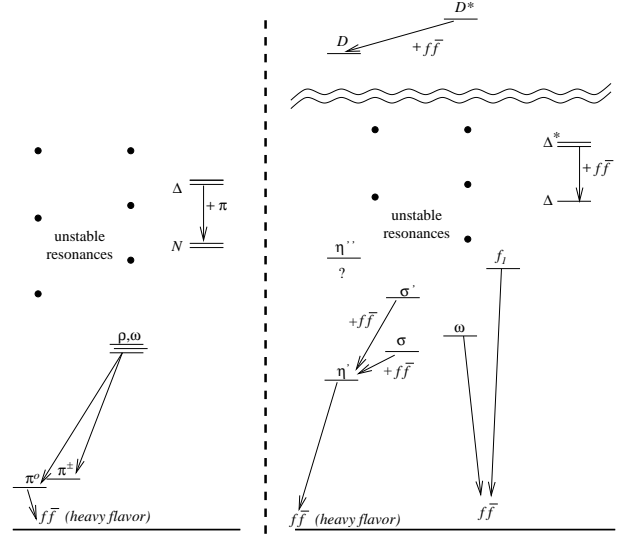}
  \end{center}
%  \vskip -0.25in  
\caption{Partial
spectrum and decay modes in the two-light-flavor regime (left)
and one-light-flavor regime (right); the latter is partly guesswork.}
  \label{fig:12LFspect}
\end{figure}

\begin{figure}[htbp]
  \begin{center}
    \leavevmode
     \epsfxsize=2.7in
     \hskip 0in \epsfbox{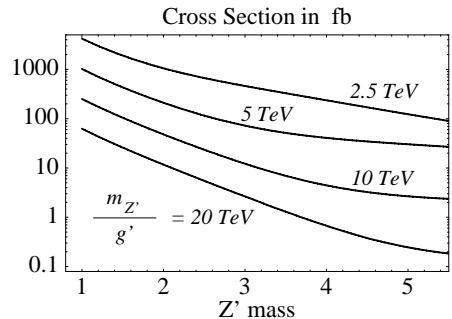}
  \end{center}
  \vskip -0.25in  \caption{Tree-level 
v-quark production cross-section in fb 
versus $m_{Z'}$, for different $m_{Z'}/g'$; 
here $\checker n=3$, $k=0$, $q_+=-2$, 
and $\checker\Lambda,m_U,m_C\ll m_{Z'}$.}
  \label{fig:LHCcs}
\end{figure}

\nonoindent\underline{Two Light Flavors (2LF):} We first
consider the regime $m_U\sim m_C\ll \checker
\Lambda$, where the physics 
resembles that of QCD (Fig.~\ref{fig:12LFspect}).
An approximate v-isospin between $U$ and $C$
quarks controls the spectrum; by analogy we call the 
(neutral!) v-pions $\checker\pi^\pm, \checker\pi^0$.
All v-hadrons decay rapidly to v-pions 
and v-nucleons, of which all are stable 
and invisible
except the $\checker\pi^0$. The $\checker\pi^0$
is unique: with v-quark wave function $U\bar U-C\bar C$,
it can decay via $Q\bar Q\to Z'\to f\bar f$, where
$Q=U,C$ and $f$ is any SM fermion.  
%(There is no
%v-photon and no analogue of
%$\checker\pi^0\to\gamma\gamma$ decay.)  
Using
$Q_f+Q_{\bar f}=\pm Q_H$ for all $f$, and $q_++q_-=-Q_\phi$,
we find (except near $m_{\checker\pi}\sim m_Z$, 
where $\Gamma_Z$ must be included),
\begin{equation}\label{Gpi2ff}
\Gamma_{\checker\pi^0}
= {1\over 8\pi}{g'^4 \over m_Z'^4}{
Q_\phi^2Q_{H}^2 f_{\checker\pi}^2 m_{\checker\pi}^5
\over (m_{\checker\pi}^2-m_Z^2)^2} 
\sum_f N_{c}^{f} m_f^2 v_f \ .
\end{equation}
Here $m_f$, $v_f$ are the mass and velocity
of $f$; $N_{c}^{f}=3$ for quarks (1 for leptons). 
(For very small $m_{\checker\pi}$ or
$Q_H$, a loop effect may be dominant.)  
The extra factor of
$(m_{\checker\pi}/m_Z)^4$ at low $m_{\checker\pi}$
arises because the axial currents for
$Q$ and $f$ decouple at low momentum.
Heavy flavor is favored,
due to the same helicity-flip that enhances $\mu^+ \nu$  in
usual $\pi^+$ decay.   If
$2m_b<m_{\checker \pi}<2m_t$, the dominant decay is $\checker \pi\to b\bar b$.
For $m_{\checker\pi}\gg m_Z$
\begin{equation}\label{Gpitothi}
\Gamma_{\checker\pi\to b\bar b} \sim
3\times 10^{15} {\rm \ sec}^{-1}{f_{\checker\pi}^2 m_{\checker\pi}
\over (200 \GeV)^3}
\left({10 \TeV\over  m_{Z'}/g'}\right)^4 \ ,
\end{equation}
while for $m_{\checker\pi}\ll m_Z$
\begin{equation}\label{Gpitotlo}
\Gamma_{\checker\pi\to b\bar b} \sim
6\times 10^{9} {\rm \ sec}^{-1}{f_{\checker\pi}^2 m_{\checker\pi}^5\over 
(20 \GeV)^7}
\left({10 \TeV\over  m_{Z'}/g'}\right)^4  \ . 
\end{equation}
(Here and elsewhere, we show
explicit numbers  for the simple case
$\checker n=3$, $k=0$, $q_+=-2$.)  
We will see (Fig.~\ref{fig:LHCcs}) 
that $m_{Z'}/g'$ can range widely, as can $m_{\checker\pi}$ and 
$f_{\checker\pi}$.
Consequently $\checker\pi^0$ decay may be
prompt, or may occur with a displaced vertex
in any element of the detector.

\nonoindent\underline{One Light Flavor (1LF):} 
The regime $m_C\gg \checker \Lambda
\gg m_U$ is unfamiliar; its  spectrum and decay
modes are not precisely known.  
Multiple v-hadrons are stable against v-strong decays.
For any $\checker n$, these include both the lightest pseudoscalar
$\checker\eta'$ and the lightest vector $\checker\omega$ (if C is
conserved.)  The axial anomaly ensures
$m_{\checker\eta'}/m_{\checker\omega}\propto \sqrt{1/ \checker n}$
\cite{etaprimemass}.  Therefore, for small $\checker n$, there will be
other narrow resonances, including the scalar $\checker\sigma$ (which
is wide in QCD,) the first excited v-baryons $\checker\Delta^*$,
and the vector $C\bar U$ meson (which we call $\checker D^*$
by analogy with charm mesons.)
For $\checker n=3$, approximate mass degeneracies
inherited from supersymmetry, analogous to the conjecture in
\cite{orbifold}, can be rigorously demonstrated \cite{orientifold} to
be applicable \cite{oneflavor}.  It is expected \cite{predictions}
that the $m_{\checker\eta'}$ and 
$m_{\checker\sigma}$  differ by no more than 30 percent; 
the $\checker\omega$
is similarly accompanied by a $J^P=0^+$ state $\checker\sigma'$, while the
$f_{1v}$ has a partner $0^-$ state $\checker\eta''$.  Other states
may be unstable.

For definiteness we assume the $\checker\eta'$ is the
lightest v-hadron (Fig.~\ref{fig:12LFspect}).  The $\checker \eta'$ decay is
similar to that of the $\checker \pi^0$; the formulas 
(\ref{Gpi2ff})-(\ref{Gpitotlo}) 
apply with simple adjustments.
The $\checker\omega$ decays, via $U\bar U\to Z'\to f\bar f$, are
typically prompt; 
ignoring $m_f$ and with $m_{\checker\omega}-m_Z$ not too small,
\begin{eqnarray}
\Gamma_{\checker\omega}
&\sim& {1\over 96\pi}{g'^4\over m_{Z'}^4}
(\Delta q)^2 A^2 m_{\checker\omega}^5  
\sum_f N_c^f K\left(Q_f,Q_{\bar f},{m_{\checker\omega}\over m_Z}\right) 
\nonumber \\
&\sim &
4\times 10^{18} {\rm \ sec}^{-1}
\left({m_{\checker\omega}\over 200 \GeV}\right)^5
\left({10 \TeV\over  m_{Z'}/g'}\right)^4 \ ,
\end{eqnarray}
where in the last line we took 
$A^2 \equiv |\ME{{\bf 0}}{J^\mu_U+J^\mu_{\bar U}}{\checker\omega}|^2
= 1$ and $\Delta q=q_+-q_-=-2$.   
The function $K$ includes effects of $Z-Z'$ mixing; it
has a resonance near $m_Z$ and for large $m_{\checker\omega}$
goes to $Q_f^2+Q_{\bar f}^2$. For $k=0$
and $m_{\checker\omega}<2m_t,2m_N$,
$\sum N_c^f K \approx 6\ (9) $ 
and $Br(\checker\omega\to\mu^+\mu^-)\sim 6\%\ (3\%)$
for $m_{\checker\omega}\gg m_Z$ 
$(m_{\checker\omega}\ll m_Z)$.
(Decays to $W^+W^-$ can occur through $Z'$--$Z$ mixing;
the branching fraction, while small, 
may be experimentally interesting.)
A metastable $f_{1v}$ decays similarly.  
Other v-hadrons
such as $\checker\sigma$, $\checker\sigma'$, $\checker
\Delta^*$ and $\checker D^*$ most likely have three-body
decays to $f\bar f$ 
plus a lighter v-hadron (Fig.~\ref{fig:12LFspect}); note the
$\checker \Delta$ and $\checker D$ are stable.
These decays are 
sensitive to small mass splittings.
Due to unknown v-hadronic form factors, {\it etc.},
lifetimes can only be estimated:
\begin{eqnarray}
\Gamma&\sim& {1\over 60\pi^3}{g'^4\over m_{Z'}^4}
(\Delta m)^5
\sum N_c^f K\left(Q_f,Q_{\bar f},{\Delta m\over m_Z}\right)
\nonumber \\
&\sim&
3\times 10^{14} {\rm \ sec}^{-1}\left({\Delta m\over 50 \GeV}\right)^5
\left({10 \TeV\over  m_{Z'}/g'}\right)^4
\end{eqnarray}
where $\Delta m$ is the mass-difference between the initial
and final v-hadron.
Finally, although $\checker D$ is invisible,
$C\bar C$-onium states decay by annihilation
to light, visible v-hadrons; these decays would allow access
to $m_C$.

\underline{Unstable $C$:} In the model above, the $C$ is
stable, but many variants of this model naturally have
flavor-changing interactions that cause $C$ decays.  These could
include higher dimension operators such as $C \bar U U\bar
U$, induced at tree level (as would occur if there are two scalars
$\phi_1,\phi_2$ instead of one) or by loops (as with v-squark
masses in a supersymmetric variant of the model.)  Rather than explore
models here, we may view this as introducing one more free parameter:
the lifetime for the lightest $C\bar U$ v-meson(s).  In the 1LF
regime, the $\checker D$ and $\checker D^*$ are heavy, and v-hadronic
decays such as $\checker D\to \checker\eta'\checker \omega$,
$\checker\eta'\checker\sigma\checker\omega$, {\it etc.}, are
kinematically allowed.  This can lead to final states with
four or more jets and/or leptons, which may appear promptly or emerge
from a displaced vertex anywhere in the detector, depending on the
strength of the flavor-changing coupling.  By contrast, in the 2LF
regime, kinematics forbids v-hadronic decays of the $\checker\pi^+$.
The $\checker\pi^+$ instead
decays 
to heavy-flavor $f\bar f$ or to $f\bar f \checker \pi^0$.  Its 
lifetime is much
longer than that of the $\checker\pi^0$, Eq.~(\ref{Gpi2ff}).

\nonoindent\underline{Existing Experimental Constraints:} 
Cosmological constraints are minimal: v-strong interactions ensure
that all v-hadrons annihilate efficiently to the lightest
v-mesons, failing to do so
only if there are asymmetries in conserved quantities such as v-baryon
number or $C$-number.  (Indeed, such asymmetries could permit stable
v-mesons or v-baryons to be self-interacting dark matter.)
Were all v-mesons stable, they would form a cold gas and dominate over
ordinary matter at temperatures somewhat below the scale
$\checker\Lambda$, which would ruin big-bang nucleosynthesis (BBN); 
but as long as one
light v-meson has a lifetime $\ll 1$ second, requiring $\checker\Lambda \agt
1$ GeV, processes such as $\checker \pi^+\checker
\pi^-\to\checker\pi^0\checker\pi^0$ followed by rapid $\checker\pi^0$ decay
(and analogous processes in the 1LF regime)
reduce the density of v-pions exponentially before BBN occurs.

For models with light v-hadrons, LEPI can place strong constraints.
Results from $2\times 10^7$ $Z$ decays limit the branching fraction
for $Z\to U\bar U, C\bar C$, via $Z$--$Z'$ mixing, although the degree
of constraint depends on the dominant final states.  A rough estimate
is given by treating the v-quarks as free:
\begin{eqnarray}\label{LEPIbound}
\Gamma_{Z\to U\bar U} &\sim& {\checker n\over 6\pi}  {g'^4\over m_{Z'}^4}
Q_H^2 
(q_+^2+q_-^2){ m_Z^5\over g_1^2+g_2^2} 
\end{eqnarray}
This gives a branching fraction, if $m_{Z'}/g'=10 \TeV$ and $\checker
n=3$, of about $10^{-7}$.  (Note it will be enhanced if
$m_{\checker\omega}\sim m_Z$.)  Thus, {\it if the v-hadronic decays of
the $Z$ are easily distinguished from background}, then $m_{Z'}/g'>10
\TeV$.  Even then, LHC cross-sections for $U,C$ production for
$m_{Z'}=2.5$ TeV, $g'=0.25$, are of order 20 fb
(Fig.~\ref{fig:LHCcs}).  Moreover, many models evade these
constraints.  In the 2LF regime, the decay $Z\to
\checker\pi^0\checker\pi^0$ is forbidden, while
$Z\to\checker\pi^+\checker\pi^-$ is invisible, and the branching
fraction for the partly visible
decay $Z\to \checker\pi^+\checker\pi^- \checker\pi^0$ is of
order ten times smaller.
The 1LF regime is more complex; 
multi-object final states with $b$s have significant backgrounds, but a
widely displaced vertex could reveal them easily.  More detailed study
is needed.

LEPII constraints are weaker than the bound (\ref{LEPIbound}) but may
be important if that bound is evaded.  As each experiment recorded
$<1$ fb$^{-1}$ of integrated luminosity, and as backgrounds to
multi-jet and jets-plus-missing-momentum events are larger than at LEPI, the cross-section limit is probably weaker, and potentially much
weaker, than 1 fb.  The $e^+e^-\to Q\bar Q$ cross-section at 200 GeV
is of order 1 fb for $\checker n=3$ and $m_{Z'}/g'\sim 6 \TeV$, though
this requires more careful treatment, especially near the v-hadron resonance
region.

\nonoindent\underline{Production and Signals at Hadron Colliders:}
Given these constraints, cross-sections at the LHC are typically in
the 1-100 fb range, and 1 fb or less at Tevatron, unless the LEP
constraints are weaker due to kinematics or unobservable signals.
Production of v-quark pairs at the LHC will occur at
and below the $Z'$ resonance.  Since $q\bar q\to U\bar U$ is analogous
to $e^+e^-\to q\bar q$, the production of v-hadrons recapitulates
the physics at ADONE, SPEAR, VEPP and PETRA.  
At these experiments,
pion multiplicities
(for $E_{cm}\gg m_{\rho}$) averaged roughly $2\ln E/m_\pi$, with
considerable spread \cite{multiplicity}.  Events were spherical
at $E\alt 3$ GeV and became increasingly two-jet-like as
$E$ increased \cite{asphericity}.  
Thus LHC v-quark production will lead to
multiple v-hadrons, with the ratio $m_{Z'}/\checker\Lambda$ 
determining the average multiplicity and
whether the v-hadrons are distributed spherically,
in two collimated v-jets, or something in between.

\begin{figure}[htbp]
  \begin{center}
    \leavevmode
     \epsfxsize=.35 \textwidth
     \hskip 0in \epsfbox{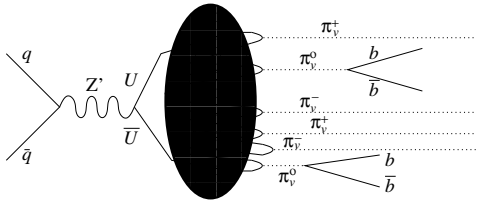}
  \end{center}
  \vskip -0.25in  \caption{A possible event
in the two-light-flavor regime; note $\pi_v^\pm$ is electrically
neutral and invisible.}
  \label{fig:2LF}
\end{figure}

For the 2LF regime, the dominant signal is multiple $b$-jet pairs
(from $\checker \pi^0$ decays) plus missing energy (from invisible
v-hadrons), with a wide distribution in the number of $b$ jets and of
the total missing energy (as in Fig.~\ref{fig:2LF}).  The missing energy,
muons from $b$ quarks, and occasional
$\tau$ pairs may assist with triggering and event selection.  If the
$\checker\pi^0$ is light, the jet pairs may be soft, may often merge,
or be otherwise hard to identify, making triggering and event
selection subtle; displaced vertices may assist in any discovery.  If
the $\checker\pi^0$ is heavy, the jets will be harder, but the number
of $b$ pairs may be smaller, decays will be prompt, and both QCD and
$Z$ plus jets will be irreducible backgrounds.  However, in this case
LEP constraints are completely evaded and production cross-sections
could be much larger.

\begin{figure}[htbp]
  \begin{center}
    \leavevmode
     \epsfxsize=.35 \textwidth
     \hskip 0in \epsfbox{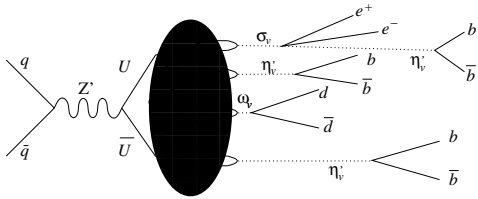}
  \end{center}
  \vskip -0.25in  \caption{A possible event
in the one-light-flavor regime.}
  \label{fig:1LF}
\end{figure}

The 1LF regime, with a greater variety of v-hadrons and final states,
produces more lepton pairs.
The $\checker\omega$ will appear as a $\ell^+\ell^-$ resonance; this
will be drowned in Drell-Yan background unless events are required to
have many $b$s or an unusual displaced vertex.  Especially interesting
are the final states from $\checker\sigma$ and $\checker\sigma'$,
where $f\bar f$ pair emission is followed by an $\checker\eta'$ decay.
One may observe $\sigma'\to \mu^+\mu^-b\bar b$, with $m_{b\bar
b}=m_{\checker\eta'}$ and $m_{\mu^+\mu^-}<
m_{\checker\sigma'}-m_{\checker\eta'}$.  Even more spectacular decays
are possible, with several objects emanating from a displaced vertex
(or two), if the $C$ is unstable.  One challenge is that lepton
isolation may be subtle here; another is that displaced vertices may
appear in the beampipe, the tracker, or even the calorimeter.
Moreover, a given event may produce several v-hadrons, which can
combine these distinctive signals into a busy and unusual event (as in
Fig.~\ref{fig:1LF}), in which identification of jets may be
challenging.  As in the 2LF regime, large $\checker\Lambda$ means
fewer v-hadrons per event and fewer displaced vertices, but any jets
and leptons are harder, and the LEP constraints on the total
cross-section are weaker or absent.

Let us add a few assorted remarks.
\newline
$\bullet$ The model we have
chosen is a bit pessimistic \cite{Langacker}, in that the $Z'$ has large
couplings to leptons (increasing LEP constraints) and small couplings to
$u$ quarks (reducing Tevatron and LHC production rates.) \
The constraints quoted here are conservative; many models can have
larger cross-sections at both LHC and the Tevatron.
\newline
$\bullet$
The high multiplicity of v-hadrons, especially
for small $\checker\Lambda$, has
many implications.  In events where
most particles decay in the detector, jets
and isolated leptons can be difficult to identify.  Conversely, in
models where the average v-hadron decays promptly, or in models
with most decays 
outside the detector,
the multiplicity 
increases the odds of seeing a straggler that decays with a visible
vertex.
It also enhances the possibility of detecting decays
of long-lived v-hadrons (as might occur for unstable $C$, and 
would occur in the model mentioned below)
by instrumenting the detector hall or a nearby cavern.
\newline
$\bullet$
Given (\ref{LEPIbound}),
a GigaZ machine would likely be able to observe $Z$ decays
to light v-hadrons.  A high-luminosity $e^+e^-$ collider 
could also study vector v-meson resonances
and their v-hadronic decay products.
\newline
$\bullet$
There has been interest in Higgs decays to multiple scalars, which
in turn decay to heavy-flavor pairs \cite{NMSSM,WeirdHiggs}.
Our v-model may initially mimic this scenario, 
since the 2LF and 1LF regimes both have a (possibly light)
pseudoscalar with $Br(\checker \pi,\checker\eta' \to f\bar f) \sim m_f^2$.
\newline
$\bullet$ We have taken the $N_i$ to be stable, but this need
not be the case.  Their striking decays in usual $Z'$ models \cite{Langacker}
would be further augmented by v-hadronic final states.

\nonoindent\underline{Higgs Mixing:} Potentially of great importance
is the effect of the mixing of $H$ with $\phi$, via a $|H|^2|\phi|^2$
coupling, along the same lines as \cite{Wells}.  This allows $gg\to
h\to Q\bar Q$, which is unaffected by LEP constraints and can
potentially increase the v-hadron production rate at the LHC and
especially at the Tevatron.  Decay modes for some v-hadrons may be
affected.  Kinematics permitting, the Higgs can even decay to
v-hadrons.  Though rare, these exotic Higgs decays could be so
distinctive, if they have displaced vertices and/or leptons, as to
possibly allow the Tevatron to discover the Higgs with its present
data.  This requires asking the right analysis questions, such as
\cite{mumu}, though in a more systematic and comprehensive fashion.
The masses, mixings and branching fractions of the $H$ and $\phi$ are
very model-dependent; a separate study of these phenomena will be required.

\nonoindent\underline{Other Models:} Other regimes of this theory,
models with more v-quarks, and other $Z'$ models will typically have
similar phenomenology but differ in important details.  (For instance,
if $m_U, m_C\gg \checker\Lambda$, the many stable glueballs will have
longer lifetimes; many will decay outside the detector, and displaced
vertices will be common.)  $Z'$ models with supersymmetry, a little
higgs, extra dimensions, {\it etc.}, would have additional diverse
v-phenomenology that we will not discuss here.  Instead, to provide a
wider perspective, we conclude with a class of models that generate
qualitatively different phenomena.

Consider adding to the SM an $SU(\checker n)$ gauge group, and
particles $X,\bar X$ charged under both color and $SU(\checker n)$,
with $m_X\sim 0.5-3$ TeV.  The v-spectrum includes
several metastable v-glueballs of mass $\sim \checker\Lambda$
and various spins \cite{Morningstar}.
A loop of $X$ particles induces
dimension-eight operators including ${\cal O}_8\equiv {\tr G^2}\ {\tr \checker G^2}$,
where $G$ ($\checker G$) is the field strengths for gluons $g$ (v-gluons
$\checker g$.)  
All 
v-glueball states can decay through these operators;
those that
decay to $gg$ have lifetimes of order
%
%\begin{equation} 
$\sim 8 \pi M_X^8/  
\alpha_s^2\checker\alpha^2 {\checker\Lambda^9}
$.
%\end{equation}
%
which for $m_X\sim 1 \TeV$, $\Lambda\sim 100$ GeV are in the range
1 psec--1 nsec.  
Glueballs decaying to $ggg$ live at least $10^3$
times longer.  
Heavy-quark and lepton pairs are absent, making 
identification of the signal challenging.  
However, given the range of glueball lifetimes, 
displaced vertices are not unlikely.
Note there are no constraints from LEP; 
decay of cosmic confining strings
may be a cosmological problem, but can be evaded if
necessary.

While ${\cal O}_8$ allows v-glueballs to be produced
through $gg\to \checker g\checker g$, $gg\to g \checker
g\checker g$, {\it etc.}, 
pair production of $X\bar X$ also suffices
and has a large
cross section.  The phenomenology of these events 
can vary widely.  Consider three possibilities, among
many others:
\begin{figure}[htbp]
  \begin{center}
    \leavevmode
     \epsfxsize=.28 \textwidth
     \hskip 0in \epsfbox{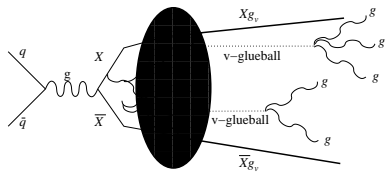}
  \end{center}
\caption{$X\bar X$ pair production with
associated v-glueballs, for a stable $X$ in the adjoint of $G_v$;
the $Xg_v$ bound state resembles a stable gluino.
}
  \label{fig:XXbar}
\end{figure}
\newline
$\bullet$
If $X$ is an adjoint of $SU(\checker n)$ and 
decays immediately to a v-gluon plus one (or more) SM quarks or gluons,
then each $X$ will decay to one (or more) hard jets
recoiling against v-glueballs (each giving softer jets).
\newline
$\bullet$
If $X$ is a stable triplet (or
octet) of color and an adjoint of $SU(\checker n)$, it will
bind to a v-gluon, forming a composite color-triplet (or
octet.)  This composite $(X\checker g)$ 
can be distinguished from a stable
quark (or gluino) by the multiple v-glueballs that will often
accompany it, {\it e.g.} 
in $q\bar q\to X\bar X\checker g$ events (Fig.~\ref{fig:XXbar}).
\newline
$\bullet$
If $X$ is stable and in the $\checker n$ of $SU(\checker n)$,
then $X$ and $\bar X$ will be bound by a v-confining
flux tube.
The bound system will lose energy by gluon emission (giving a soft
jet) or emission of v-glueballs (giving multiple jets); 
eventually $X$ and $\bar X$ will
annihilate to gluons (giving two or more ultra-hard jets) or to
v-gluons (generating
several v-glueballs.)

\nonoindent \underline{Final Remarks:} Clearly the range of possible
v-phenomena is wide.  Can it be observed?  Whether backgrounds are
large depends strongly on parameters.  In some cases, trigger
efficiencies may be an issue; in almost all cases, event selection for
an analysis is a complex matter.  High-multiplicity events with many
soft jets are especially challenging; defining jets and isolated
leptons may be problematic.  The phenomenology is rich and complex and
will require novel simulation techniques.  We hope this work will
stimulate discussion of how to ensure that physics of this type does
not go undetected.

We thank P.~Langcker, H.~Lubatti, A.~Nelson, S.~Sharpe, M.~Shifman,
P.~Skands and G.~Watts for helpful assistance.  MJS thanks T.~Han,
P.~Langacker and K.~Freese for discussions of related topics in 2001.
Some of these ideas were presented at the Atlas Muon Workshop at
U. Washington, July 2004.  This work was supported by U.S. Department
of Energy grants DE-FG03-00ER41132 (KZ) and DE-FG02-96ER40956 (MJS).

\end{document}